\documentclass[preprint,showpacs,preprintnumbers,amsmath,amssymb]{revtex4}
\usepackage{graphicx}% Include figure files
\usepackage{dcolumn}% Align table columns on decimal point
\usepackage{bm}% bold math
\begin{document}
\bibliographystyle{apsrev}

%\preprint{}

\title{Classical and quantum Coulomb crystals}% Force line breaks with \\

\author{M.~Bonitz}
\email{bonitz@physik.uni-kiel.de}
\author{P.~Ludwig}
\author{H.~Baumgartner}
\author{C.~Henning}
\author{A.~Filinov}
\affiliation{Institut f\"ur Theoretische Physik und Astrophysik, Christian-Albrechts-Universit\"{a}t zu Kiel,
D-24098 Kiel, Germany}

\author{D.~Block}
\author{O.~Arp}
\author{A.~Piel}
\affiliation{Institut f\"ur Experimentelle und Angewandte Physik, Christian-Albrechts-Universit\"{a}t zu Kiel,
D-24098 Kiel, Germany}

\author{S.~K\"ading}
\author{Y.~Ivanov}
\author{A.~Melzer}
\author{H.~Fehske}
\affiliation{Institut f\"ur Physik, Ernst-Moritz-Arndt-Universit\"{a}t, D-17487 Greifswald, Germany}

\author{V.~Filinov}
\affiliation{Institute for High Energy Density, Russian Academy of Sciences, 
Izhorskaya 13/19, 127412 Moscow, Russia}

\date{\today}

\begin{abstract}
Strong correlation effects in classical and quantum plasmas are discussed. In particular,
Coulomb (Wigner) crystallization phenomena are reviewed focusing on one-component non-neutral plasmas in traps and on macroscopic two-component neutral plasmas. The conditions for crystal formation in terms of critical values of the coupling parameters and the distance fluctuations and the phase diagram of Coulomb crystals are discussed.
\end{abstract}
\maketitle

\section{Introduction}\label{into_s}
Coulomb crystals (CC) -- a periodic arrangement of charged particles --
are omnipresent in nature, from astrophysical systems (interior of dwarf stars, Refs. \cite{segretain,chabrier93}) to laboratory systems (trapped ions, see e.g. Refs.~\cite{itano,Drewsen}, plasmas in storage rings, e.g. Refs.~\cite{schaetz01,schramm01} or dusty plasmas, Refs.~\cite{chu94,thomas94,hayashi94}, to name a few examples, for an overview see Ref.~\cite{dubin}). CC add an interesting new species to the large family of crystals in condensed matter, chemistry or biology, 
for an (incomplete) list, see table \ref{tab1}. 
We will distinguish CC from ``traditional'' crystals (including molecular or ion crystals or metals) by 1. the {\em governing role of the Coulomb interaction} (in contrast 
to crystals of neutral particles) and 2. by the {\em elementary character of the constituents} (in contrast e.g. to the complex ions forming the lattice of a metal) \cite{trappedcc}. These two properties bring the CC into the area of plasma physics rather than condensed matter physics, because it is the strength and long range of the Coulomb interaction which dominates the many-particle behavior in these systems, the crystal symmetry, stability and melting properties.

The research on CC originates in solid state physics. More than seven decades ago Wigner predicted, using the jellium model, that electrons in metals would form, at low density, a bcc lattice, see Ref.~\cite{wigner}.
A second line of research grew out of the field of classical strongly coupled plasmas. There it was predicted, by computer simulations, that a one-component Coulomb or Yukawa model plasma (OCP) in three and two dimensions would crystallize at sufficiently high density and/or low temperature, e.g. Ref.~\cite{dubin}. 3D Coulomb crystals show a bcc symmetry whereas Yukawa crystals have a bcc and a fcc phase, Ref.~\cite{hamaguchi}. In contrast, the ground state of 2D crystals has hexagonal symmetry. However, jellium and OCP are models assuming that the charge species forming the crystal coexists with a second neutralizing one which forms a static homogeneous background which does not influence the crystal. Such systems do not exist in nature. In real two-component plasmas crystallization is very different. One important effect is weakening of the Coulomb interaction by dynamic screening. Moreover, the attractive force between different species will favor recombination, i.e. formation of bound states. This will, obviously, strongly reduce the Coulomb coupling and may even prevent crystal formation. Nevertheless, CC formation in a two-component plasma (item AIII.~(c) in table \ref{tab1}) is possible and will be discussed below in Sec. \ref{2cp_s}. 

But before that we consider the second possibility to achieve Coulomb crystallization:  one-component (non-neutral) plasmas which are stabilized by an external ``trap'', such as an electric potential, cf. item BII in the table. This principle has been successfully used in experiments with ion crystals, e.g. Refs.~\cite{itano,Drewsen} and dusty plasmas, e.g. 
Refs.~\cite{melzer94,Goree96,hayashi99,arp04,antonova06}, for an overview see Refs.~\cite{piel02,fortov05}, and is expected to function also with electrons in semiconductor quantum dots, Ref.~\cite{afilinov-etal.prl01}. Naturally, the existence of the trap may have a strong influence on the crystal properties. For example, a spherically symmetric trap will favor crystals forming concentric rings (in 2D) or shells (in 3D). This gives rise to interesting symmetry effects, including magic (closed shell) configurations, e.g. Refs.~\cite{bedanov,ludwig-etal.05pre,bonitz-etal.prl06,totsuji05} familiar from atoms and nuclei and coexistence of shells and bulk behavior in larger systems, Ref.~\cite{schiffer02}.

\begin{table}\label{tab1}
\caption{Coulomb crystals (CC) in the world of crystals (incomplete list). CC variants are A.III.b, A.III.~(c), B.II.~(a) and B.II.~(b).
1CS (2CS) denotes one (two) component systems, OCP - the one-component plasma model containing 
ions plus a homogeneous static neutralizing electron background.
}
\begin{tabular}[t]{ll}
\hline\hline
{\underline{\bf A. Unconfined (macroscopic) crystals}}  & {\underline{\bf B. Confined crystals (1, 2 or 3D traps)}}\\ [0.3ex]
&\\ [0.2ex]
{\em I. 1CS with attractive interactions}                  & {\em I. 1CS with attractive interactions}  \\ [0.2ex]
neutral particles (e.g. Lennard-Jones,    & confinement not necessary, see A I.   \\ [0.2ex]
Morse potentials)                         &    \\ [0.2ex]
``normal'' solids, rare gas clusters etc. &    \\ [0.2ex]
&\\ [0.2ex]
{\em II. 1CS with repulsive interactions} & {\em II. 1CS with repulsive interactions}    \\ [0.2ex]
a. transient ``Coulomb exploding'' crystals        &   a. classical: ions, dust particles   \\ [0.2ex]
b. charges on surfaces of finite systems           &   b. quantum: electrons in quantum dots   \\ [0.2ex]
(e.g. electrons on helium droplets)                &      \\ [0.2ex]
&\\ [0.2ex]
{\em III. 2CS}                                           & {\em III. Periodic confinement}  \\ [0.2ex]
a. ``normal'' crystals: ionic crystals, metals etc.   & e.g. particles in optical lattices  \\ [0.2ex]
b. OCP model (ion Coulomb or Yukawa crystal)          & electrons in bilayers, superlattices etc. \\ [0.2ex]
c. TCP crystals (electrons, nuclei, holes, positrons) &  \\ [0.2ex]
\hline\hline
\end{tabular}
\end{table}

Coulomb crystals may not only consist of classical ``point particles'' but also of quantum particles which have a finite extension (electrons in quantum dots, ions in compact stars etc.) which is of relevance for the properties of CC and is crucial for the phase diagram. Since the issue of quantum plasmas has come into the focus of recent research again in the context of laser plasmas \cite{kremp99} and astrophysics \cite{shukla} we will consider the influence of quantum effects in some detail.
In this paper we study some general properties of Coulomb crystals. Starting from the theoretical 
description, in Sec. \ref{model_s}, we continue with two typical examples of classical and quantum crystals in traps 
(Secs. \ref{ccc_s}, \ref{qcc_s}). This is followed by an analysis of the melting point, Sec. \ref{crystal_condition_s}, after which the special situation of CC in neutral plasmas (Sec. \ref{2cp_s}) is discussed.

\section{Model and parameters}\label{model_s}
The Hamiltonian of a system of particles with mass $m_i$ and 
charge $e_i$ interacting via a statically screened Coulomb (a Yukawa) potential is given by
\begin{eqnarray}
\hat H &=& \sum\limits_{i=1}^{N} \left[ -\frac{\hbar^2}{2 m_i} \nabla^2_i
+ V({\bf r}_i) + 
\sum\limits_{j < i}^{N} \frac{e_i e_j}{\epsilon}  
\frac{e^{-\kappa r_{ij}} } {r_{ij}} \right]\,,
\label{3D_Ham}
\end{eqnarray}
where $r_{ij}=|{\bf r_i}- {\bf r_j}|$ and $\epsilon$ denotes a static background dielectric constant, which is of the order of $10$ in case of an electron-hole plasma in a semiconductor; in a plasma, $\epsilon=1$. The case of a pure Coulomb system follows in the limit 
of zero screening, $\kappa \rightarrow 0$. In the case of trapped systems a confinement potential $V(r)$
is included which will be assumed isotropic and parabolic, i.e. $V(r)=m\omega^2r^2/2$. The limit of an 
unconfined system is achieved by letting $\omega \rightarrow 0$.
In thermodynamic equilibrium the system properties are determined by the canonical probability distribution $P$ or, in the quantum case, by the density operator ${\hat \rho}$
\begin{eqnarray}
P(E)  = \frac{1}{Z}e^{-\beta E}, \qquad
\hat \rho = \frac{1}{Z}  e^{-\beta {\hat H}},
\label{rho}
\end{eqnarray}
where $\beta=1/k_BT$ is the inverse temperature, and $E$ denotes the total potential energy, 
i.e. the 2nd plus 3rd term of Eq. (\ref{3D_Ham}).

Despite their different form of appearance, all Coulomb (Yukawa) systems exhibit similar
fundamental properties governed by the strength 
of the Coulomb (Yukawa) interaction which is measured by dimensionless control parameters: the coupling parameters 
$\Gamma_a$, $r_{sa}$ and $\lambda_a$ of particle species ``a'' and the quantum degeneracy parameter $\chi_a$. These parameters are determined by the ratio of characteristic 
energy and length scales~\cite{bonitz-book,bonitz-etal.03jpa}: 
\begin{itemize}
\item{
{\em Length scales}: (i) ${\bar r}$ -- average inter-particle distance,
${\bar r} \sim n^{-1/d}$ ($n$ and $d=1,2,3$ denote the density and
dimensionality of the system, respectively).
(ii) $\Lambda$ -- quantum-mechanical extension of the particles. For free
particles we have $\Lambda^{free}_a=h/\sqrt{2\pi m_a k_B T_a}$
(DeBro\-glie wavelength), for bound particles $\Lambda$ is given by
the extension of the ground state wave function, $\Lambda^{bound}_a=2\pi a_B$.
(iii) $a_B$ -- relevant Bohr radius 
$a_B=\frac{\epsilon}{e_ae_b}\frac{\hbar^2}{m_{ab}}$,
where $m^{-1}_{ab}=m^{-1}_{a}+m^{-1}_{b}$. (iv) $a_{Ba}$ -- effective Bohr radius of an OCP: 
$a_{Ba}=\frac{\epsilon}{e^2_a}\frac{\hbar^2}{m_{a}}$.
}
\item{
{\em Energy scales}: (i) $\langle K \rangle$ -- mean kinetic energy, 
which in a classical system is given by 
$\langle K_a \rangle_{cl} = \frac{d}{2} k_BT_a$,
whereas in a highly degenerate Fermi system 
$\langle K_a \rangle_{qm}= \frac{3}{5} E_{Fa}$ holds
[$E_F=\hbar^2(3\pi^2 n)^{2/3}/2m$ denotes the Fermi energy];
(ii) $\langle U^{ab}_c \rangle$ -- mean Coulomb energy, given for free 
and bound particles by 
$\langle U^{ab}_c \rangle_{f}=\frac{e_ae_b}{4\pi \epsilon} \frac{1}{\bar r}$ and  
$\langle U^{ab}_c \rangle_{B}=\frac{e_ae_b}{4\pi \epsilon}\frac{1}{2
a_B}\equiv E_R$ (Rydberg), respectively. Analogously, the mean Yukawa interaction energy 
is estimated by $\langle U_Y \rangle_{f}=e^{-\kappa \bar r}\:\langle U_c \rangle_{f}$.
}
\item{
{\em Dimensionless control parameters}:
The quantum {\em degeneracy parameter} 
$\chi_a \equiv n_a\Lambda_a^d\sim (\Lambda_a/{\bar r}_a)^d$
divides many-body systems into classical ($\chi < 1$) and quantum
mechanical ones ($\chi\ge 1$).
The {\em Coulomb coupling parameter} is the ratio
$|\langle U_c\rangle|/\langle K\rangle$. For classical systems 
 $\Gamma_a\equiv |\langle U^{aa}_c\rangle|/k_BT_a$ results,
whereas for quantum systems the role of $\Gamma_a$  is taken over 
by the Brueckner parameter, $r_{sa}\equiv {\bar r}_a/a_{Ba} \sim |\langle U^{aa}_c\rangle|/E_{Fa}$. The relation 
to the parameter, $r_s={\bar r}/a_{B}$, familiar from atomic units is $r_s=r_{sa}[1-\frac{m_a}{m_a+m_b}]$.
Similarly one can introduce coupling parameters for Yukawa systems and of different species.
}
\end{itemize} 

In a {\em two-component plasma} different masses and charges of the species may give rise to unequal  coupling and quantum degeneracy of the species. In particular, in a dense electron-ion plasma classical ions and quantum electrons may coexist. Analogously ions maybe strongly coupled while the electrons are only weakly coupled, see Sec. \ref{2cp_s}.
The ratio of the degeneracy parameters scales as 
$\chi_a/\chi_b=(m_b/m_a)^{1/2}$,  whereas the ratios of the coupling parameters are given by
$\Gamma_a/\Gamma_b=(e_a/e_b)^{2-1/d}$ and $r_{sa}/r_{sb}=(m_a/m_b)(e_a/e_b)^{2+1/d}$, where local charge neutrality, $n_ae_a=n_be_b$, has been assumed.

\section{Classical Coulomb and Yukawa crystals in traps}\label{ccc_s}

\begin{figure}[h]
\begin{center}
\includegraphics[height=4.5cm,clip=true]{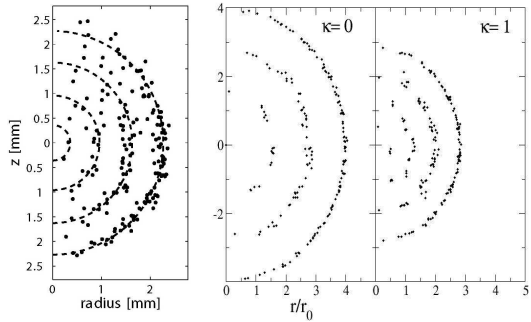}
\end{center}
\vspace{-0.3cm}
\caption{Radial particle distribution for $N=190$ particles given in cylindrical coordinates. Left: experiment, right two figures: simulation results with Coulomb ($\kappa=0$), 
and Yukawa ($\kappa=1$) potential. The length unit in the right two figures is $r_{oc}$,
given by Eq. (\ref{eq:r0}), from Ref. \cite{bonitz-etal.prl06}.}
\label{yuball_config}
\end{figure}

\begin{figure}[ht]
\begin{center}
\begin{tabular}[t]{|c|c|c|c|c|}
\hline
$\kappa$   & $N_1$     & $N_2$   & $N_3$ & $N_4$\\% [0.2ex]
\hline
\hline
$0$ & $1$    & $18$   & $56$   & $115$ \\ 
$0.2$ & $1$   & $18$  & $57$  & $114$ \\ 
%$0.3$ & $2$   & $20$  & $57$  & $111$ \\ 
$0.4$ & $2$  & $20$ & $58$ & $110$\\ 
%$0.5$ & $2$  & $21$ & $58$ & $109$\\ 
$0.6$ & $2$  & $21$ & $60$ & $107$\\ 
$1.0$ & $4$  & $24$ & $60$ & $102$\\ 
\hline
Exp & $2$  & $21$ & $60$ & $107$\\ 
\hline
\end{tabular}
\label{tab:ns_190}\hspace{1.5cm}
\includegraphics[width=6cm,angle=-90,clip=true]{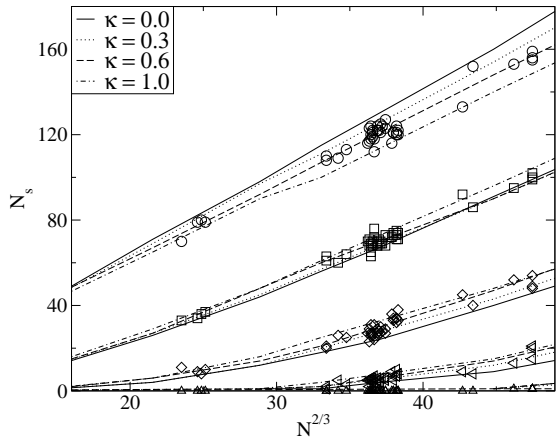}
\end{center}
\caption{
Number of particles $N_s$ on the shells of Yukawa balls with different $N$ and $\kappa$. 
Table contains experimental (last line) and theoretical shell configurations for $N=190$. $N_1\dots N_4$ denote the particle numbers on the i-th shell beginning in the center. Figure shows the shell populations for $40$ experimentally observed Yukawa balls (symbols) and molecular dynamics simulation results for several $\kappa$ values \cite{bonitz-etal.prl06}. $\kappa$ is given in units of $r_0^{-1}$ defined by $m\omega^2r_0^2=e^2/r_0$, temperature is in units of $E_0=e^2/r_0$. 
}
\label{3dcryst}
\end{figure}

Coulomb crystallization in a spherical 3D geometry was first observed for ultra-cold ions 
in Penning or Paul traps~\cite{itano}. 
A second candidate are ions created by ionization of cooled trapped atoms.  
Recent simulations~\cite{pohl04} show that the expanding ions 
might crystallize if they are properly laser cooled during the expansion. 
Finally, so-called ``Yukawa balls'' have been observed in dusty 
plasmas~\cite{arp04,block-etal.ppcf07}, see Fig. \ref{yuball_config}. Their theoretical description is again based on the Hamiltonian~(\ref{3D_Ham}) (for an overview 
on earlier theoretical results and simulations see~\cite{dubin}). 
This model has, in fact been shown to correctly describe 
the dusty plasma measurements~\cite{bonitz-etal.prl06,baumgartner-etal.cpp07}: 
3D concentric shells with the populations $N_s$ being sensitive to the screening strength 
$\kappa$. With increasing $\kappa$ the reduced repulsion leads to an increased population of the inner shells, cf. the table.
The quality of the experiments is so high that the shell populations can be measured accurately, allowing for comparisons with 
the simulations. In fact, very good agreement is found for $\kappa r_0\approx 0.6$, cf. Fig. \ref{3dcryst}, which shows the relevance of screening effects in these confined dusty plasma crystals. Furthermore, screening has an important effect on the average radial density profile of these crystals. In contrast to Coulomb crystals, where the density is approximately constant, with increasing $\kappa$ there is an increasingly rapid decay of the density towards the surface \cite{henning-etal.pre06,henning-etal.pre07}.

As in the 2D case closed shell configurations and 
a ``Mendeleyev table'' exist (see, e.g., 
Refs.~\cite{tsuruta93,ludwig-etal.05pre,bonitz-etal.prl06}). 
The dependence of the crystal stability on the
number of particles can be seen from their melting temperatures. 
For example, the closure of the first spherical shell occurs at
$N=12$, which gives rise to a particularly high 
crystal stability (high melting temperature), cf. Fig.~\ref{wcphase}.

\section{Quantum Coulomb crystals in traps}\label{qcc_s}

\begin{figure}[h]
\begin{center}
\includegraphics[width=10cm,clip=true]{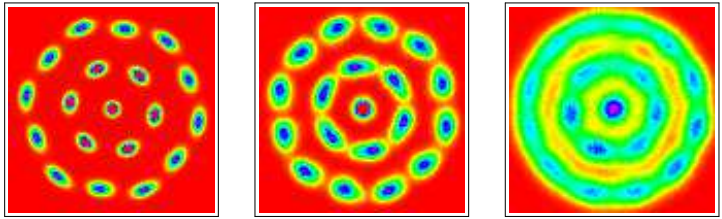}
\end{center}
\vspace{-0.3cm}
\caption{(Color) 19-electron quantum Wigner ``crystal'' (left), radially ordered crystal (center) and mesoscopic fermionic liquid (right). From left to right quantum melting at constant temperature occurs. Dots correspond to the 
probability density $\rho$ of the electrons 
in the 2D plane which varies between 
$\rho_{max}$ (pink) and zero (red).}
\label{ecryst}
\end{figure}

When the trapped CC is cooled, eventually the DeBroglie wavelength $\Lambda$ will exceed the interparticle distance and quantum effects will become relevant. While for ion crystals this may require sub-microkelvin temperatures this regime is easily accessible with (the much lighter) electrons in nanostructures. At the same time, there quantum crystal formation and detection is hampered by impurities and defects. Therefore, the results shown below are obtained by means of computer simulations. The density operator (\ref{rho}) with the 2D Hamiltonian (\ref{3D_Ham}) is evaluated by performing first-principle path integral Monte Carlo (PIMC) simulations, for details see Refs.~\cite{filinov-etal.ppcf01,comp_met06}. Results for the probability density of 19 electrons in a 2D harmonic trap are shown in Fig. \ref{ecryst}. We observe a shell structure similar as in the classical case. However, the particles are now not point-like but have a finite extension and an elliptic shape which minimizes the total energy. When the system is compressed by increasing $\omega$, the wave functions of the electrons start to overlap -- first within each shell, cf. central part of Fig. \ref{ecryst}, and finally also particles on different shells overlap giving rise to a quantum liquid state. This process of {\em quantum melting} occurs even at zero temperature, giving rise to an interesting phase diagram of quantum CC \cite{afilinov-etal.prl01}, see also Sec. \ref{crystal_condition_s}.

\section{Conditions for crystal formation}\label{crystal_condition_s}

Phase coexistence is determined by the equality of the thermodynamic potentials (such as the free energy) in the two phases which often requires very accurate and expensive calculations. In macroscopic plasmas there exist many alternative criteria for crystallization: peaks of the specific heat, sufficiently strong modulations of the pair distribution or the static structure factor and so on. 
These quantities yield practically the same melting point, for an analysis see e.g. Ref.~\cite{hartmann}. In contrast, in trapped systems, in particular, when the particle number is reduced, the results for the phase boundaries may strongly depend upon the chosen quantity and the way it is computed. It turns out that, for the class of systems described by Eq.~(\ref{3D_Ham}) two quantities are particularly useful to localize the melting point: critical values of the coupling parameter and of the distance fluctuations of the particles around their equilibrium positions. We mention that, for very small systems, recently a more appropriate quantity has been proposed - the variance of the block-averaged interparticle distance fluctuations, see Ref.~\cite{boening-etal.08prl}.

\subsection{Critical values of the coupling parameter}\label{crit_coup_ss}
Let us start with the simplest case of  Eq. (\ref{3D_Ham}) -- a {\em macroscopic classical plasma} ($\omega=0$) containing a single charge component. We can rewrite the ratio of energy and temperature which determines the thermodynamic properties, cf. (\ref{rho}), as 
\begin{eqnarray}
\beta E = f(\Gamma,\kappa) =
\Gamma\sum\limits_{1\le j < i}^{N} \frac{e^{-{\bar \kappa} {\bar r}_{ij}} } {{\bar r}_{ij}} \quad 
\mbox{with} \quad {\bar r}_{ij}=\frac{r_{ij}}{{\bar r}}, \quad {\bar \kappa}=\kappa {\bar r}.
\label{be1c}
\end{eqnarray}
For Coulomb systems ($\kappa=0$) $\beta E$ is characterized by a single parameter, the coupling parameter $\Gamma$, i.e. different Coulomb systems (containing different types of particles, having different temperature or density) are expected to show the same behavior if they have the same values of $\Gamma$. In particular, as was revealed by simulations, CC occurs at $\Gamma_{cr}\simeq 175$ in 3D and $\Gamma_{cr} \simeq 137$ in 2D. In a Yukawa OCP ($\kappa>0$) the effect of screening suggests to introduce $\Gamma_{Y}(\kappa) \rightarrow \Gamma e^{-\kappa {\bar r}}$, however, this does not correctly reproduce the $\kappa-$dependence of the melting curve. The reason is that melting is not determined by the absolute value of the energy but by the energy contribution of particle fluctuations around their ground state positions $r_{i0}$. Expanding (\ref{be1c}) around 
$r_{i0}$, defining $\xi_{ij}={\bar r}_{ij}-{\bar r}_{ij0}$ and taking into account that the first derivatives vanish we obtain
$\beta \Delta E = \beta (E - E_0 - E_{com}) = \Gamma\sum_{j < i}^{N} 
\frac{\xi_{ij}^2} {{\bar r}^3_{ij0}}\left(1+{\bar \kappa}{\bar r}_{ij0} + 
\frac{{\bar \kappa}^2{\bar r}_{ij0}^2}{2}\right)e^{-{\bar \kappa} {\bar r}_{ij0}}  + \dots$. The dots denote terms with mixed derivatives and higher order terms, and $E_0$ and $E_{com}$ are the energy in the ground state and of center of mass excitations (which are not relevant for the melting), respectively.

For the case of two particles, this expression can be written in a Coulomb-like form,  
$\beta {\bar r}_0^3\Delta E  = \Gamma_{Y}(\kappa)\xi^2$, with the Yukawa coupling parameter
$\Gamma_{Y}(\kappa)=\Gamma e^{-\kappa {\bar r}}[1+\kappa {\bar r}+(\kappa {\bar r})^2/2]$. 
Assuming that, at the melting point, the critical coupling parameter is universal (3D case), 
$\Gamma_{Ycr}=175$, the phase boundary of the (bcc) crystal in the $\Gamma-\kappa$ plane 
is approximated by $\Gamma_{cr}(\kappa)=175 \cdot e^{\kappa {\bar r}}[1+\kappa {\bar r}+(\kappa {\bar r})^2/2]^{-1}$. Interestingly, simulations have shown that this result holds reasonably well not just for small particle numbers but also in a macroscopic system \cite{fortov03}.

\begin{figure}[t]
\begin{minipage}{1.0\textwidth}
\begin{center}
\vspace{-0.0cm}
\mbox{\includegraphics[width=7.2cm,clip=true]{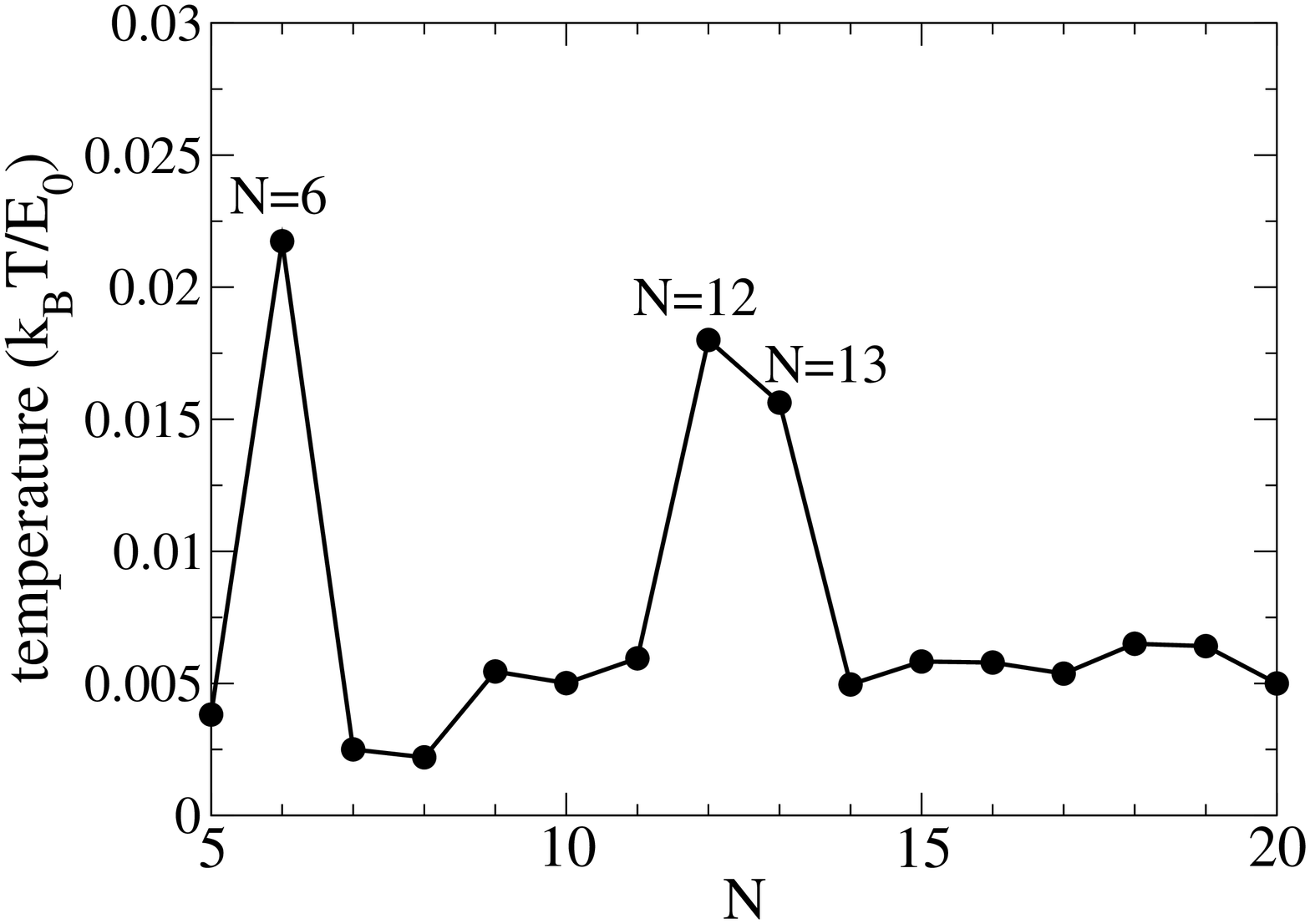}}
\vspace{-0.1cm}
\mbox{\includegraphics[width=6.5cm,clip=true]{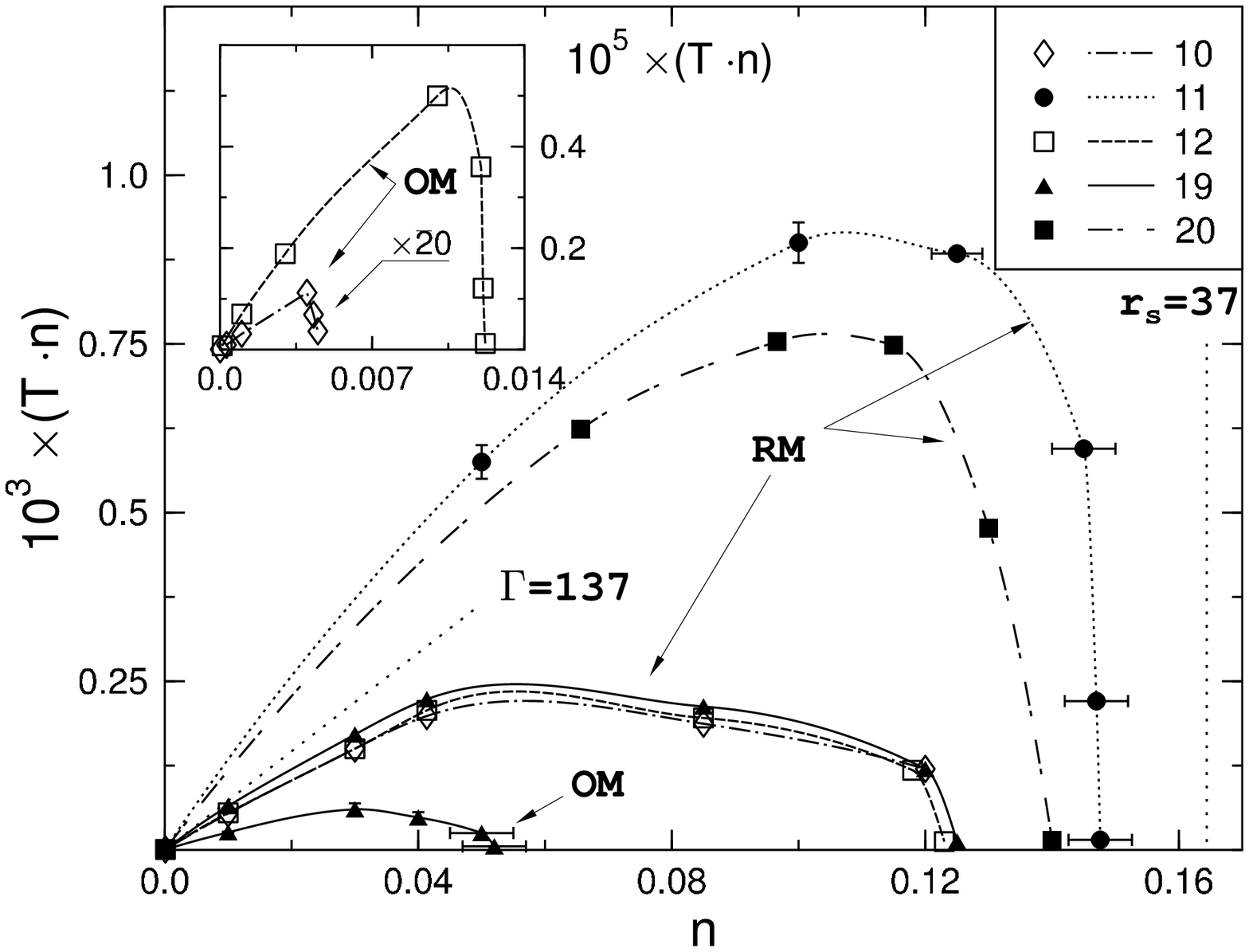}}
\end{center}
%\vspace{-1.5cm}
\end{minipage}% Dies Prozent ist wichtig! (kein horiz. Abst. zw. minipages)
\caption{Left: Melting temperature of small 3D spherical Yukawa crystals versus particle number. 
Right: Phase diagram of the mesoscopic 2D Wigner crystal for different particle numbers $N$. OM (RM) denotes the boundary of orientational (radial) melting. Here the dimensionless density $n$ 
and temperature $T$ are defined as $n= \sqrt{2} \,l_0^{2}/r^2_0=(a_B/r_0)^{1/2}\approx r_s^{-1/2}$ and $T= k_B T/E_0$, where $l^2_0 = \hbar/(m\omega_0), E_0 = e^2/\epsilon_b {r_0}$ with 
$r_0$ given by $e^2/\epsilon_b {r_0}=m\omega^2 r^2_0/2$, from Refs. \cite{vova-etal.jpa06,afilinov-etal.prl01}.}
\label{wcphase}
\end{figure}

Consider now a {\em classical crystal in a trap}. Here the density is externally controlled by the trap frequency $\omega$ which determines the mean interparticle distance. The basic properties are best illustrated for two particles. The ground state is obtained from the minimum of the relative potential energy (\ref{3D_Ham}) with the result 
\begin{eqnarray}
\frac{e^{\kappa r_0}r_0^3}{1+\kappa r_0} = \frac{e^2}{\frac{m}{2}\omega^2} 
\equiv r^3_{0c}.
\label{eq:r0}
\end{eqnarray} 
Eq.~(\ref{eq:r0}) yields the two-particle distance, $r_{0}(\kappa)$, as a function of the distance in an unscreened system, $r_{0c}$ \cite{bonitz-etal.prl06}. In analogy to the macroscopic case we introduce a Coulomb coupling parameter, $\Gamma_2\equiv e^2/(k_BT r_0)$. The corresponding coupling parameter for Yukawa interaction, $\Gamma_{2Y}$, again follows from expansion of the energy around the ground state,  
$\beta r_0^3\Delta E  = \Gamma_{2Y}(\kappa)\xi^2$ with the result  \cite{bonitz-etal.prl06} $\Gamma_{2Y}=\Gamma_2 e^{-\kappa r_0}\left( 1+\kappa r_0+\kappa^2r_0^2/3 \right)$ slightly differing from the above expression.
In a similar way, the ground state and effective coupling parameter can be defined for any particle number, but this has to be done numerically \cite{ludwig-etal.05pre,arp-etal.jp05}. The results are strongly $N-$dependent due to the importance of shell filling and finite size effects. This leads to strong variations of the crystal stability with $N$ as can be seen in the melting temperatures, left part of Fig. \ref{wcphase}, see. e.g. Refs.~\cite{vova-etal.jpa06,peeters07pre}.

Consider now a {\em macroscopic quantum OCP}. We rewrite the 
Hamiltonian (\ref{3D_Ham}) in dimensionless units 
\begin{eqnarray}
\beta \frac{\hat H}{2E_R} = g(r_s, T, \kappa) = - \frac{\beta}{r_s^2}\sum_i \nabla^2_{{\bar r}_i}
+\frac{\beta}{r_s}\sum\limits_{1\le j < i}^{N} \frac{e^{-{\bar \kappa} {\bar r}_{ij}} } {{\bar r}_{ij}}, \label{be1q}
\end{eqnarray}
which depends on the quantum coupling parameter $r_s$ and temperature separately, leading to a more complex behavior than in a classical OCP where only one parameter $\Gamma$ exists. The existence of three energy scales - quantum kinetic energy (first term), interaction energy (second) and thermal energy has a direct consequence for the phase boundary $T_{cr}(n)$ of Coulomb crystals, cf. Fig. \ref{wcphase}. While for a classical crystal, the slope of the boundary is always positive, ${\rm d}T_{cr}(n)/{\rm d}n>0$, given by a constant value of $\Gamma$, for quantum crystals, 
there exists a maximum value of the temperature, $T^{max}_{cr}$, where the slope changes sign. For densities to the left of the maximum the phase boundary is dominated by ``normal'', i.e. thermal melting, whereas for  densities exceeding the value of the maximum, by a competition of quantum kinetic and interaction energy. For sufficiently large densities (with decreasing $r_s$) quantum melting is observed, even at zero temperature, cf. Fig.~\ref{wcphase}. The corresponding critical values of the Brueckner parameter of a Coulomb OCP at $T=0$ are $r^{cr}_{s}\approx 100 (160)$ in 3D and  $r^{cr}_{s}\approx 37$ in 2D for fermions (bosons) \cite{ceperley80,afilinov-etal.prl01} and references therein. These values are still under investigation. Also, generalization of the results to a quantum Yukawa OCP has only recently been attempted, see Ref.~\cite{militzer} and refences therein. 

{\em Finite trapped quantum plasmas} show the same general behavior as a macroscopic quantum OCP and, in addition, finite size effects as in case of the classical crystals in traps. As a consequence, the crystal phase boundary is strongly $N-$dependent, as can be seen for the 2D case  in Fig. \ref{wcphase}. 
Further, in 2D the competition of hexagonal (bulk) symmetry and spherical symmetry induced by the trap leads to  possible additional phases, both in classical and quantum trapped plasmas. The most prominent one is a partially ordered phase where particle ordering occurs within each shell, but no order of different shells with respect to each other exists. Only at significantly larger values of the coupling parameter the orientational fluctuations freeze out (orientational freezing or melting, ``OM'') and the crystal enters the fully ordered phase, cf. Fig. \ref{wcphase}. The location of this phase boundary is strongly dependent on the crystal symmetry and may vary with $N$ by many orders of magnitude \cite{afilinov-etal.prl01}. In 3D trapped plasmas no radially melted phase is observed because there is generally a much larger energy barrier for intershell rotations.

\subsection{Critical values of the distance fluctuations}\label{crit_ur_ss}
The appearance of different coupling parameters in the case of classical and quantum plasmas makes it 
very difficult to construct a joint phase diagram of Coulomb crystals. An alternative approach to the crystal phase boundary uses, as the starting point, the magnitude of the relative interparticle distance fluctuations of the particles around their lattice positions. Expanding, as in Section \ref{crit_coup_ss}, the total energy fluctuations $\Delta E$ in a Taylor series up to second order and diagonalizing the result allows to express $\Delta E$ as a superposition of $d\cdot (N-2)$ relative normal modes. For this system of independent 1D quantum harmonic oscillators with the phonon modes $\omega_{\lambda}(q)$ of polarization $\lambda$ all thermodynamic properties at a given temperature $T$ are known. For example, the thermodynamic average of the distance fluctuations $\langle \delta x^2 \rangle = \langle x^2 \rangle - \langle x \rangle^2$ is given by \cite{chabrier93}
\begin{equation}
\langle \delta x^2 \rangle = \frac{1}{2}\sum_q\sum_{\lambda=1}^d \frac{\hbar}{m\omega_{\lambda}(q)}f_{\lambda}(q,T),
\quad \mbox{with} \; f_{\lambda}(q,T)=\coth{\frac{\hbar \omega_{\lambda}(q)}{2k_BT}}.
\label{hl_dx} 
\end{equation}
For a macroscopic classical OCP, $f_{\lambda}(q,T) \rightarrow \frac{2k_BT}{\hbar \omega_{\lambda}(q)}$, and the average over the phonon spectrum yields, in case of a bcc crystal, 
$\langle \delta x^2 \rangle = 12.973 \, {\bar r}^2/\Gamma$. The result for the relative distance fluctuations $u_{rel}\equiv \sqrt{\langle \delta x^2 \rangle/r_0^2}$  normalized to the nearest neighbor distance, $r_0 = (3\pi^2)^{1/6}{\bar r}$, is
\begin{equation}
u^{cl}_{rel}=\sqrt{ \frac{12.973\,}{(3\pi^2)^{1/6}} \frac{1}{\Gamma} } \longrightarrow 0.155,
\label{ucl} 
\end{equation}
where the last number is the critical value obtained by using $\Gamma=175$.

Analogously, we obtain for a quantum OCP bcc crystal at zero temperature, where $f_{\lambda}(q,T) \rightarrow 1$, 
\begin{equation}
\langle \delta x^2 \rangle = \frac{3^{1/2}}{2}u_{-1}r^{3/2}_{sa}a^2_{Ba},
\label{dxq} 
\end{equation}
with $u_{-1}\equiv \langle \frac{\omega_{pa}}{\omega_{\lambda}} \rangle$ denoting the moment of order minus one of the phonon spectrum which equals $2.7986$ for a bcc crystal \cite{baiko00}. This yields, for the relative distance fluctuations
\begin{equation}
u^{q}_{rel}=\sqrt{\frac{0.783}{r^{1/2}_s}} \rightarrow 0.28 \; (0.249),
\label{uq} 
\end{equation}
where the last number is the critical value for fermions (bosons), using $r^{cr}_s=100$ (160). Note that these fluctuations are mainly due to quantum diffraction effects, i.e. the finite extension of the particle wave functions. Spin effects (quantum exchange) play a minor role for the location of the crystal phase boundary which is clear since, in the crystal state, the wave function overlap has to be small. Nevertheless, the physical properties of crystals of bosons maybe essentially different from the one of fermions. The reason is that interacting bosons may show superfluid behavior which may even persist in the crystal phase. This state is called a {\em supersolid} and was predicted thirty years ago \cite{chester67,andreev69,legget70} and was recently observed in PIMC simulations of trapped bosonic plasma crystals \cite{afilinov-etal.08}.

Equations (\ref{ucl}) and (\ref{uq}) are very useful as they establish the relation between relative distance fluctuations and the relevant coupling parameter in the two limiting cases of classical and quantum plasmas. To connect the two limits along the whole phase boundary, cf. Fig. \ref{wcphase},
one has to use the full phonon spectrum, Eq. (\ref{hl_dx}), without expansion of the function $f_{\lambda}$.
The temperature and density dependence of $u_{rel}$ remains an open question although some interpolations have been attempted, see e.g. Ref.~\cite{chabrier93}. Further improvements, in particular, in the quantum regime, 
may require to include anharmonic corrections, e.g. Ref.~\cite{dubin90}, since their finite extension lets the particles explore ranges of the potential energy which cannot be approximated by a parabola, e.g. Ref.~\cite{militzer}.

\section{Unconfined two-component Coulomb crystals}\label{2cp_s}

As discussed in the introduction, crystal formation in two-component plasmas (TCP) competes with bound state formation. One may, therefore, ask whether there exist parameters where CC exist and, at the same time, Coulomb bound states are ionized.
In comparison to an OCP, in a TCP, we have at our disposal two additional parameters to realize these two conditions: 
the mass ratio $M=m_h/m_e$ and charge ratio $Z=e_h/e_e$ (in a non-equilibrium mass-asymmetric plasma there is further the possibility of different temperatures of the components \cite{bonitz-etal.prl05}). 
The first requirement is obvious: the heavy component (ions or holes) has to be sufficiently strongly correlated such that it can form an
{\em OCP} Wigner crystal. The second condition is that electrons have sufficiently high kinetic energy to escape the ionic binding potential. For classical electrons this requires a sufficiently high temperature whereas in a quantum plasma ionization is possible when electron wave functions of neighboring atoms start to overlap -- this leads to tunnel ionization (Mott effect) which occurs at a sufficiently high density. In summary, we find two alternative sets of conditions
\begin{eqnarray}
\Gamma_i  & \ge & \Gamma^{cr} \qquad \mbox{and}  \quad  \frac{d}{2}k_BT_e > E_R, \qquad {\rm classical} \; {\rm case},
\label{cc}
\\
r_{si} & \ge & r_s^{cr} \qquad \mbox{and} \quad r_{se} < r_{se}^{\rm Mott}, \qquad{\rm quantum} \; {\rm case},
\label{qc}
\end{eqnarray}
where in 3D $r_{se}^{\rm Mott}\approx 1.2$.
The phase boundary of the Coulomb crystal of the heavy particles can be obtained using the harmonic lattice theory results of Sec. \ref{crit_ur_ss}. For the quantum case, we may use, for $\langle \delta x_h^2 \rangle$, Eq.~(\ref{dxq}) and express the nearest neighbor distance of the heavy particles, $r_{0h}$, by the one of the electrons
\begin{equation}
u^2_{rel,h}=\frac{\langle \delta x_h^2 \rangle}{r^2_{0h}}=
\frac{3^{1/2}u_{-1}}{2\alpha^2}\frac{r^{3/2}_{sh}}{Z^{2/3}r^{2}_{se}}\frac{a^2_{Bh}}{a^2_B},
\label{uqh} 
\end{equation}
where $\alpha_{bcc}=(3\pi^2)^{1/3}$. Assuming that, at the phase boundary, the critical value of the fluctuations is given by the OCP result, Eq.~(\ref{uq}), and $r_{se}=r_{se}^{\rm Mott}$ we readily obtain the existence conditions of a CC of fermionic (bosonic) ions in a two-component plasma: $M^{cr}Z^{4/3}= 83.3 \; (132.8)$. 
This agrees with the result of Ref. \cite{bonitz-etal.prl05} where it was obtained from a different derivation. Thus crystallization requires a minimum mass ratio $M$ between heavy and light particles. This condition is fulfilled for compact dwarf stars where a crystal of carbon and oxygen nuclei (fully ionized atoms) is expected to exist \cite{segretain,chabrier93}. Further candidates are crystals of protons which was recently confirmed by PIMC simulations, cf. Refs.~\cite{filinov-etal.jetpl00,militzer}, or  $\alpha-$particles, see Ref.~\cite{bonitz-etal.prl05}. Both systems might be accessible in laboratory experiments in the near future. Another area where such two-component CC should be observable are electron-hole plasmas in intermediate valence semiconductors, see Refs.~\cite{bonitz-etal.prl05,bonitz-etal.jpa06} where one could also verify the critical value of $M$ experimentally although values of $M$ as large $80$ exist only in some special materials. Another promising candidate are charge asymmetric bilayers where hole crystallization is expected to occur already for $M \lesssim 10$ \cite{ludwig-etal.cpp07} which is due to the 2D confinement of the particles.
\begin{figure}[h]
\begin{center}
\includegraphics[width=12.cm,clip=true]{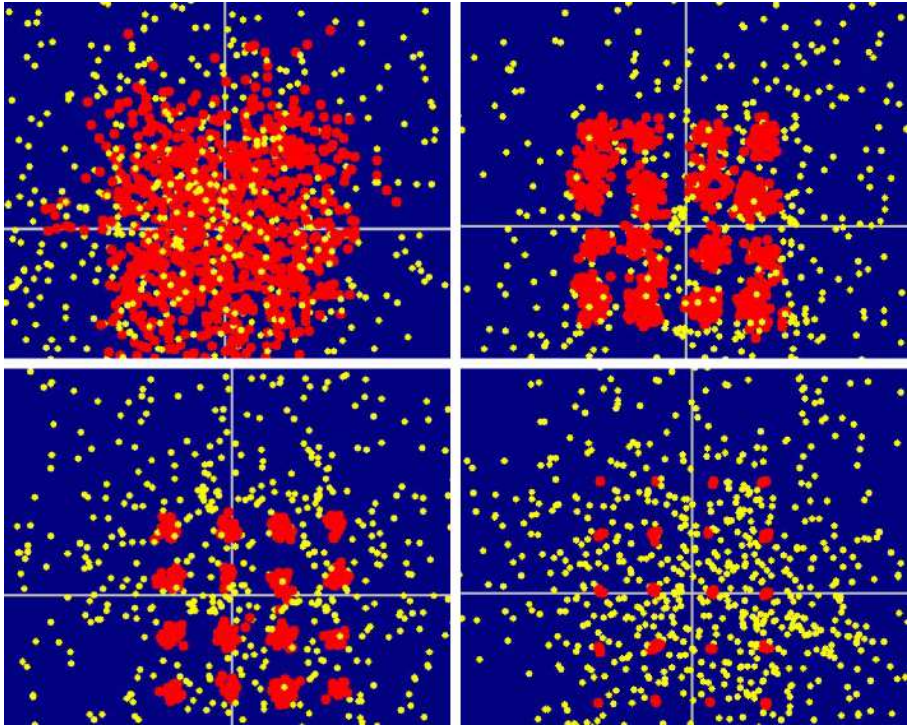}
\end{center}
%\vspace{-1.5cm}
\caption{\small (Color) Snapshots of a Coulomb crystal of heavy particles (red clouds) embedded into a Fermi gas of electrons (yellow) in a macroscopic two-component (neutral) plasma (spin averaged results) for mass ratio $M=12$ (top left), $M=50$ (top right), $M=100$ (bottom left), $M=400$ (bottom right). The density corresponds to $r_{se}=0.64$, the temperature is $T_e=T_h=0.06 E_R$. First-principle two-component PIMC simulations.}
\label{2csnaps}
\end{figure}
   \hspace{-1.3cm}

\begin{figure}[htp]
%\begin{center}
%\hspace*{-8cm}
\includegraphics[width=16.cm,clip=true]{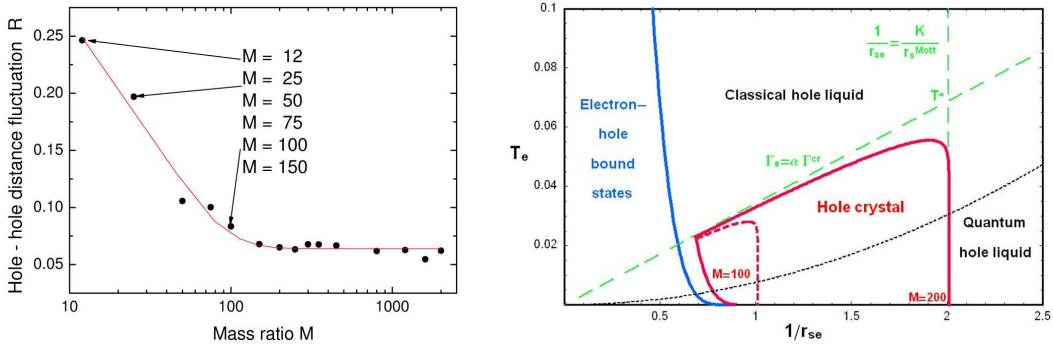}
%\includegraphics[width=5.cm,angle=-90,clip=true]{gr5.eps}
%\makebox(0,0){\put(-0,-270){\includegraphics[width=7.cm,clip=true]{phase5.eps}}}   %Boxgröße 0,0 (unsichtbar), x,y Positon
%\end{center}
\vspace{-0.1cm}
\caption{(Color) Left: mean-square relative heavy particle distance fluctuations versus mass ratio $M$ for $T_e=0.096$ and $r_{se}=0.63$. 
Symbols are simulation results, the line is the best fit \cite{bonitz-etal.jpa06}. Right: Qualitative phase diagram of a Coulomb crystal of heavy 
particles (``holes'') in a macroscopic two-component (neutral) plasma. $T_e=\frac{3}{2}k_BT/E_R$. 
Taken from Refs. \cite{bonitz-etal.prl05,bonitz-etal.jpa06}.}
\label{2cphase}
\end{figure}

The analytical predictions of heavy particle crystallization in a TCP have been verified by PIMC simulations where both electrons and heavy particles have been treated fully quantum mechanically \cite{bonitz-etal.prl05,bonitz-etal.jpa06,filinov-etal.pre07}. As can be seen in Fig.~\ref{2csnaps}, with increasing $M$, indeed hole localization becomes more pronounced and, between $M=50$ and $M=100$, a transition to crystal-like behavior is observed. A quantitative analysis based on the relative distance fluctuations of the heavy particles, left part of Fig. \ref{2cphase}, confirms that the liquid-solid transition takes place 
around $M\sim 80$. This is a novel kind of quantum phase transition, where melting occurs at constant temperature and density -- by ``changing'' the heavy particle mass.
The phase diagram of the two-component CC is sketched in Fig.~\ref{2cphase} for the two values $M=100$ and $M=200$. The larger $M$ the more extended is the crystal phase in the density-temperature plane. The crystal phase is bounded from above by the (green dashed) line $\Gamma_h=\Gamma_{cr}$ and from the right (high densities) by the
(vertical green dashed) line $r_{sh}=r_s^{cr}$. This is the simplest approximation where the influence of the electrons on the heavy particle interaction has been neglected. Improvements require the inclusion of screening effects \cite{militzer}, as discussed above, this leads to a destabilization of the crystal. At the same time, the heavy particle crystal also influences the spatial distribution of the electrons which stabilizes the crystal compared to the OCP case. Thus, there exist two competing effects for the crystal stability. A detailed comparison of the crystal phase diagram in an OCP and a TCP, therefore, remains an interesting still open question. Finally, it has been predicted by Abrikosov \cite{abrikosov78} that, in the presence of a hole crystal, the electrons should tend to form Cooper pairs, i.e. exhibit superconductivity which yet remains to be verified experimentally.

\section{Conclusions}\label{conclusions_s}
In this paper we have given an overview on strong correlation effects in classical and quantum plasmas, in particular on
Coulomb (Wigner) crystallization. We have discussed the possible occurences of Coulomb crystals, first, in trapped one-component plasmas and, second, in two-component neutral plasmas. The conditions for crystal formation have been summarized in terms of known critical values for the coupling parameters as well as in terms of critical values of the relative interparticle distance fluctuations. Using the data for the critical parameters it is possible to construct the phase diagram of strongly coupled Coulomb matter which was discussed for two cases: mesoscopic classical and quantum plasmas in a parabolic 2D trap and two-component mass-asymmetric plasmas.

\begin{acknowledgments}
This work is supported by the Deutsche Forschungsgemeinschaft via SFB-TR 24 grants 
A3, A5 and A7.
\end{acknowledgments}


\begin{thebibliography}{26}
\bibitem{trappedcc} In trapped one-component CC such as ion or dust crystals the second restriction does not apply, but these systems are clearly separated from the traditional crystals by the external confinement.
\end{thebibliography}
\end{document}